\documentstyle [aps,epsf]{revtex}

\def\figwidth{4in}

\begin{document}


\title{Spin Dynamics of the LAGEOS Satellite
       in Support of a Measurement of the
       Earth's Gravitomagnetism}

\author{Salman Habib}
\address{Theoretical Astrophysics Group (T-6, MS B288), Theoretical Division,
Los Alamos, National Laboratory, Los Alamos, NM 87545}
\author{Daniel E. Holz}
\address{Department of Physics, Enrico Fermi Institute, University of Chicago,
Chicago, IL 60637; and
Theoretical Astrophysics Group (T-6, MS B288), Theoretical Division,
Los Alamos, National Laboratory, Los Alamos, NM 87545}
\author{Arkady Kheyfets}
\address{Department of Mathematics, North Carolina State University, Raleigh,
NC  27695-8205}
\author{Richard A. Matzner}
\address{Center for Relativity and Physics Department, The University of Texas
at Austin, Austin, TX 78712-1081;
and Applied Research Laboratories,  The University of Texas at Austin, P.O. Box
8029, Austin, TX 78713-8209}
\author{Warner A. Miller}
\address{Theoretical Astrophysics Group (T-6, MS B288), Theoretical Division,
Los Alamos, National Laboratory, Los Alamos, NM 87545; and Astrodynamics
Branch (PL/VTA), Phillips Laboratory (AFMC), Kirtland AFB, NM 87117}
\author{Brian W. Tolman}
\address{Applied Research Laboratories, The University of Texas at Austin, P.O.
Box 8029, Austin, TX 78713}
\date{May 25, 1994}

\maketitle

\begin{abstract}
{\small LAGEOS is an accurately-tracked, dense spherical satellite
covered with 426 retroreflectors. The tracking accuracy is such as to
yield a medium term (years to decades) inertial reference frame
determined via relatively inexpensive observations. This frame is used
as an adjunct to the more difficult and data intensive VLBI absolute
frame measurements. There is a substantial secular precession of the
satellite's line of nodes consistent with the classical, Newtonian
precession due to the non-sphericity of the earth. Ciufolini has
suggested the launch of an identical satellite (LAGEOS-3) into an
orbit supplementary to that of LAGEOS-1: LAGEOS-3 would then
experience an equal and opposite classical precession to that of
LAGEOS-1. Besides providing a more accurate real-time measurement of
the earth's length of day and polar wobble, this paired-satellite
experiment would provide the first direct measurement of the general
relativistic frame-dragging effect. Of the five dominant error sources
in this experiment, the largest one involves surface forces on the
satellite, and their consequent impact on the orbital nodal
precession.  The surface forces are a function of the spin dynamics of
the satellite.  Consequently, we undertake here a theoretical effort
to model the spin ndynamics of LAGEOS.  In this paper we present our
preliminary results.}
\end{abstract}

\pacs{PACS numbers: 04.80.+z, 04.20.-q.41.10.Fs}

\section{The LAGEOS-3 Mission.}
\label{I}
The Laser Geodynamic Satellite Experiment (LAGEOS-3) is a joint USAF,
NASA, and ASI proposed program to measure, for the first time, a
quasi-stationary property of the earth -- its gravitational magnetic
dipole moment (gravitomagnetism) as predicted by Einstein's theory of
general relativity. This gravitomagnetic field causes local
inertial frames to be dragged around with the earth at a rate
proportional to the angular momentum of the earth, and inversely
proportional to the cube of the distance from the center of the earth.
Thus the line of nodes of the orbital plane of LAGEOS-3
precesses eastward at $32\ mas/yr$.  Although in this example the
frame dragging effect is small compared to the torque on the orbital
plane due to the oblateness of the earth, it is an essential
ingredient in the dynamics of accretion disks, binary systems, and
other astrophysical phenomena~\cite{ASTRO}.

Today, almost eighty years after Einstein introduced his geometric
theory of gravity, we have just begun to measure -- to verify -- his
gravitation theory.  Of no less stature than the ``tide producing''
$-M/r^2$ ``electric component'' of gravity is the inertial-frame
defining ``magnetic component'' of gravitation $-J/r^3$.  To see this
force in action: first, inject a satellite into a polar orbit about an
earth-like mass idealized as not spinning with respect to the distant
quasars. The satellite will remain in orbit in a continuous
acceleration towards the center-of-mass of the attracting body under
the influence of the Newtonian $1/r^2$ force, and its orbital plane
will remain fixed in orientation with respect to distant quasars.
Second, spin the central body, giving it angular momentum, and follow
the trajectory of the satellite. Its orbital plane will experience a
torque along the central body's rotation axis. The orbital plane will undergo
a precessional motion in the direction of the central body's rotation.
The mass in motion of the central body, or ``mass current'', produces a
dipole gravitational field -- the gravitomagnetic field. In the case
of a satellite orbiting at two earth radii, the orbital plane will
precess about the body axis of the earth at approximately $32\
mas/yr$. This is the Lense-Thirring effect \cite{LT}.

The Lense-Thirring force has never been directly measured. A
measurement of this gravitomagnetic force can be compared to the
pioneering work of Michael Faraday on the measurement of the magnetic
force between two current-carrying wires. However, the laboratory
setting for this gravity measurement will be the 4-dimensional curved
spacetime (approximately Kerr) geometry enveloping the earth. The idea
behind the LAGEOS gravity measurement is simple. Whereas the
Everitt-Fairbanks experiment (Gravity Probe-B) proposes putting a
gyroscope into polar orbit~\cite{GPB}, the Ciufolini LAGEOS-3/LAGEOS-1
experiment \cite{CIU} proposes the use of the orbital planes
themselves as a gyroscope.

\begin{figure}
\centerline{\epsfxsize=\figwidth\epsfbox{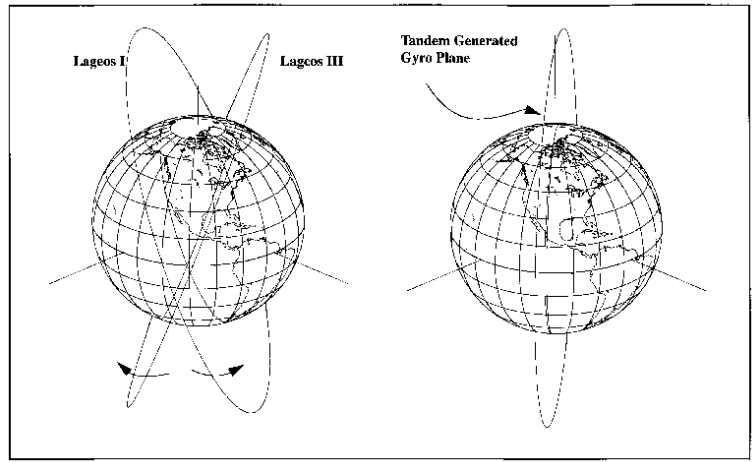}}
\caption{
{\bf The tandem-generated gyro plane and the LAGEOS-3 satellite
experiment}. Both the Gravity Probe B (GP-B) experiment at Stanford
and the LAGEOS-3 experiment share the feature of generating an
effective polar orbit so as to cancel the torque on the orbital plane
arising from the oblateness of the earth. In the case of the GP-B
experiment, a drag-free superconducting gyroscope is launched into a
polar orbit. In the case of the LAGEOS-3 experiment, a satellite is
launched into an inclination that is supplementary ($70=180-110~\deg$)
to the LAGEOS-1 satellite.  As the orbital planes precess in equal and
opposite directions, the line of intersection of these two
supplementary planes evolves so as to sweep out a polar plane.
Einstein's theory of general relativity predicts that this
tandem-generated gyroplane will precess at a rate of $32~mas/yr$.  In
the GP-B case, the gyroscope is $\sim cm$ in diameter, while it is
$12,271.79~km$ in diameter for LAGEOS.}
\label{figure1}
\end{figure}

In 1976 NASA launched the LAGEOS-1 satellite, a totally passive
$60~cm$ diameter ball of aluminum with $426$ retro-reflecting mirrors
embedded in its surface. (There are numerous globally-located laser
tracking stations to observe LAGEOS-type satellites.)  LAGEOS-1 was
injected into a two earth-radii circular orbit at an inclination of
$110~\deg$. Due to the oblateness of the earth, the orbital plane
rotates at a rate of $126~\deg/yr$. This torquing can only be modeled
to $450~mas/yr$ -- which is not accurate enough to measure the
$32~mas/yr$ gravitomagnetic force. The idea of Ciufolini is to launch
another LAGEOS satellite (LAGEOS-3) into an orbit identical to that of
LAGEOS-1, except that its inclination is supplementary ($70~\deg =
180~\deg - 110~\deg$). This proposed orbital plane will rotate in the
opposite direction, {\em i.e.}, $-126~\deg/yr$.  The intersection of
the two (LAGEOS-1, LAGEOS-3) orbital planes will sweep out a
``tandem-generated gyro plane'' (Fig.\ \ref{figure1}).  The utilization of two
satellites cancels out many of the large precessional effects due to
mass eccentricities of the earth, providing a plane inertial enough
for a measurement, accurate to five percent (or better), of the
``magnetic component'' of gravity.

\section{Why Spin Dynamics?}
\label{II}
The success of the LAGEOS experiment hinges upon the detection of a
$32~mas/yr$ eastward drift of the line of nodes of the two satellites.
A strong effort is now underway to model the orbital and spin dynamics
of these satellites, and to make an assessment of the uncertainties
these will add to the desired measurements. Encapsulated in
Table\ \ref{table1} are the five major classes of errors in this experiment:
(1) geopotential (other than even zonals) and tides, (2) earth
radiation pressure, (3) uncertainty in other relativistic effects, (4)
earth and solar-induced thermal forces, and (5) even zonal
geopotentials (per $0.1\deg$ inclination injection error). Due to the
recent GEM-T1 improvements in the earth's zonal harmonics, the errors
due to solid-earth tides have been significantly reduced, and are now
potentially smaller than those due to surface effects \cite{RIES}.

These surface effects ({\em e.g.},\ Yarkowsky thermal drag, neutral
and charged particle drag) cause a change in the nodal precession of
the satellite, thus contributing to what will be potentially the
largest source of error in the LAGEOS-3 measurement of the
gravitomagnetic effect.  In modeling these surface forces, in
particular the Yarkowsky thermal drag (also referred to as the
Rubincam effect) due to the satellite's differential heating and
delayed reradiation, the behavior of the spin vector of the satellite
is of crucial importance.  That the uncertainties induced by the
surface forces on LAGEOS are on the order of 4\%, out of a 6\%
experiment, makes such a theoretical modeling of paramount importance.
With this in mind, we undertake in this paper a theoretical model of
the spin dynamics of LAGEOS-type satellites, and compare our results
with observational data.

We must emphasize that we are looking for qualitative, not
quantitative, results. We do not intend to predict the exact magnitude
and direction of the spin vector at any particular instant. Instead,
we desire a rough estimate of its magnitude, and whether its behavior
is predictable or chaotic. We seek to answer questions of the sort:
``In the asymptotic limit, does the satellite tidally lock, tumble, or
have some other behavior?''

Previous studies of the spin dynamics of the LAGEOS-1 satellite
\cite{BI} were valid only for spin rates much larger than the orbital
frequency. Today, however, the spin of the satellite, decaying with a
three-year time constant, is rapidly approaching the orbital period.
We therefore require an analysis in the low--frequency regime. Unlike
the earlier analysis which dealt with orbit-averaged
quantities, we will solve numerically the full set of dynamical
equations. As we will show, while the low frequency regime exhibits
complex behavior, the asymptotic state of this forced, damped system
appears to be tidally locked. In this paper we will examine the
spin-orbit resonance phase and discuss the asymptotic state of the
spin of the satellite. It is the goal of this research to provide an
optimum strategy for the measurement of the spin dynamics of the
LAGEOS satellites in support of the proposed gravity measurement. In
addition, we use our theoretical model to propose an optimum orbital
injection procedure for LAGEOS-3: it is our opinion that the LAGEOS-3
satellite should be injected into the orbital plane with as large a
spin rate as possible. Our results provide the first analysis of the
asymptotic spin dynamics of the LAGEOS satellite. Previous
calculations were unable to analyze the spin-orbit resonances of the
satellite, nor its asymptotic behavior, which will play a crucial role
in the experiment.  We demonstrate in this paper that the LAGEOS-1
satellite will be sufficiently predictable to support the
gravitomagnetic measurement.

\begin{table}
\bigskip
\begin{tabular}{||l|lr||}                             \hline
Geopotential and tides                    &    2\% &\\ \hline
Earth radiation pressure                  &    1\% &\\ \hline
Uncertainty in other relativistic effects &    1\% &\\ \hline
Thermal forces                            &  1-3\% &\\ \hline
Even zonal geopotential                   & $<$1\% &\\ \hline
Random and stochastic errors              &    2\% &\\ \hline
{\bf RSS error}                           &  3-4\% &\\ \hline
\end{tabular}
\bigskip
\caption{
{\bf The latest error budget for the LAGEOS-3 Lense-Thirring
experiment}. Errors in geopotential and tides reflect improvement over
GEM-T1. Uncertainty in the thermal forces depends mainly on knowledge
of spin-axis motion. The even zonal geopotential error calculation
assumes less than $0.03~\deg$ inclination injection error. The random
and stochastic error estimate can accommodate seasonal variations in
low degree spherical harmonics of the geopotential.}
\label{table1}
\end{table}

\section{Modeling the Spin}
\label{III}
It is rather interesting that after 36 years the spin dynamics of
passive satellites is once again important to the field of
astrodynamics. In 1957 Vinti \cite{VINTI} analyzed the spin dynamics
of a non-ferromagnetic spherical satellite in the earth's magnetic
field, which was then of critical importance to the alignment of
antennas. Today we perform the same analysis on a slightly oblate
satellite, of critical importance to the first measurement of the
magnetic component of gravity, as predicted by Einstein.

There are many factors to consider when analyzing the spin dynamics of
an oblate, metallic satellite orbiting in the gravitational and magnetic
fields of the earth. The most prominent
effect is the torqueing due to the gravitational field of the earth.
This arises from the oblateness of the satellite, with the
oblateness of the earth producing a negligible contribution that can be
added to our calculations as an adiabatic correction. If the
satellite's (bulging) equatorial plane lies in the plane of orbit, no
such torques are possible. However, when the satellite is not placed
exactly in such a position, gravitational torques will arise. In an
effort to model these torques, we consider the situation of an oblate
spheroid in orbit around a point mass. As was done by Bertotti and
Iess, we parallel the development in Goldstein \cite{GOLD}.

Bertotti and Iess's analysis of the effects of gravity on the spin
dynamics of the oblate satellite, which resulted in predicting a {\em
chaotic} spin dynamics of arbitrarily large amplitude in the obliquity
at late times, is not appropriate for small rates of spin. Their
prediction is based on the ``Hipparcos'' formula for the rate of
precession $\omega_p$ of an oblate satellite in the inhomogeneous
gravity field of the earth
\begin{equation}
\omega_p = {3\over 2}\, \Delta\, {\omega_0^2\over \omega_3} \cos\theta,
\label{1}
\end{equation}
where $\omega_0$ is the orbital angular velocity, $\omega_3$ is the
satellite spin rate, $\theta$ is the obliquity angle of the satellite
(the angle between $\vec\omega$ and the normal to the orbital plane)
and
\begin{equation}
\Delta = {I_3 - I_1\over I_3} \label{2}
\end{equation}
is the satellite's oblateness. Here $I_3$ and $I_1 = I_2$ are the
principal moments of inertia (the principal direction corresponding to
$I_3$ is assumed to be that of $\vec\omega$ and coincides, by
assumption, with the symmetry axis of the oblate satellite).

It is argued by Bertotti and Iess that, since $\omega_p \propto
1/\omega_3$, the gravitational precession in the asymptotic limit of
small $\omega$ becomes very fast and may make the spin dynamics
chaotic.  This conclusion is based upon a misunderstanding. We have
shown, via a careful analysis of assumptions underlying the
``Hipparcos'' formula, that even in the approximation commonly used in
deriving the formula (averaging of the gravitational potential over
the satellite orbit, dipole cutoff of the multipole decomposition,
{\em etc.}) Eqn.~(\ref{1}) can only be used when
\begin{equation}
{6 \omega_0^2 \Delta \cos^2\theta \over \omega_3^2} \sim
{\omega_0^2 \over \omega_3^2} \ll 1,
\label{3}
\end{equation}
{\em i.e.}, when the spin rate of the satellite is much greater than
its orbital angular velocity. The latter restriction is easy to
overcome, and the corrected equation for $\omega_p$ is
\begin{equation}
\omega_p = {1\over 2} {\omega_3 \over \cos\theta}
\left( 1 - \sqrt{1 + {6 \omega_0^2 \Delta \cos^2\theta \over \omega_3^2}}
\right).
\label{4}
\end{equation}
This equation imposes a bound on $\vert\omega_p\vert$:
\begin{equation}
\vert\omega_p\vert < \omega_0 \sqrt{{3\over 2} \Delta},
\label{5}
\end{equation}
which makes it clear that the rate of precession cannot grow to
cause chaoticity of the satellite spin dynamics. A subsequent
qualitative investigation \cite{FUCHS} by Chris Fuchs has shown that when
magnetic forces are included in the picture the precession rate
remains bounded, and should be much smaller than $\omega_0$. Another
conclusion reached in the course of our analysis has been that, when
both gravitational and magnetic forces are taken into account, the
nutation, although bounded in its amplitude, does not disappear
completely even in the asymptotic limit. The value of these results,
however, is limited by the fact that they do not yield the exact
bounds, nor do they provide any information on the time
scale to reach the asymptotic limit.  However, they do clearly show
that chaoticity of the LAGEOS spin dynamics caused by an unbounded
growth of gravitationally-induced precession cannot occur.

\begin{figure}
\centerline{\epsfxsize=\figwidth\epsfbox{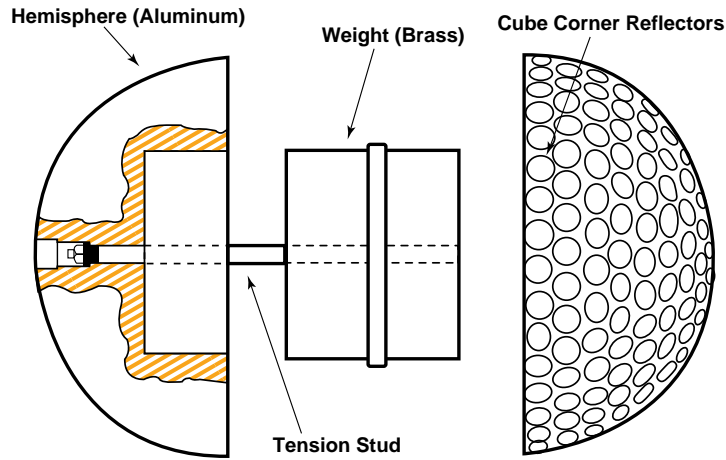}}
\caption{
{\bf A schematic representation of the LAGEOS assembly}. The two
hemispheres form a $60~cm$ diameter sphere and are made of $6061$
aluminum. There are $426$ cube corner reflectors recessed into the
satellite's surface.  To obtain the desired surface-to-mass ratio, the
inside of the aluminum hemispheres were hollowed out to make room for
a brass cylinder of $31.76~cm$ diameter and $26.70~cm$ height. The
LAGEOS-1 satellite was launched June 7, 1976 into an inclination of
$109.8585~\deg$ and semimajor orbit of $12271.790~km$. (This figure is
adapted from Ref. \protect\cite{JOHN}.)}
\label{figure2}
\end{figure}

Another important factor governing the evolution of the spin vector is
the interaction of the metallic core of the satellite with the
magnetic field of the earth. The LAGEOS satellite \cite{JOHN} is
composed of two aluminum hemispheres bolted together, with a brass
cylindrical core along its body axis (the original axis of spin)
(Fig.\ \ref{figure2}).  The spinning of this metallic object in the
magnetic dipole field of the earth (and the motion through that field)
will cause eddy currents within the satellite, which will in turn
cause dissipation through Joule heating and a slowdown of the spin,
and furthermore will cause torques on the spin vector. These torques
can be understood as the interaction of the magnetic dipole, caused by
the induced eddy currents, with the magnetic field of the earth. In
modeling this effect, we have treated a uniform, spherical object in
the orbit of a perfect magnetic dipole. Here we are concerned with a
simple qualitative analysis of the spin dynamics of LAGEOS, as the
complexity of the true satellite geometry prohibits us from a precise
model of the eddy current distribution.  Our theoretical model permits
us to analyze arbitrary inclinations of the orbital plane to the
earth's magnetic dipole axis.  The results presented in this paper
correspond to a retrograde $I=109.86~\deg$ orbit.

The problem of a spinning metallic sphere in a constant magnetic field
has been treated by Landau and Lifshitz \cite{LL}, and we avail
ourselves of their results. For our purposes, we ignore the torques
caused by the changing magnetic field due to the orbit (as opposed to
spin) of the satellite. These torques can be shown to be negligible
until asymptotically late stages of motion, and have no qualitative
effects upon the dynamics.

The source of many of the difficulties in doing analyses of such
orbiting, spinning bodies lies in the involved coordinate systems
needed to describe their motion. Thus, it is important at this point
to give a brief description of the coordinates we will use in this
paper. In our analysis of the spin dynamics of the LAGEOS satellite we
found it convenient to introduce the following four coordinate
systems:

{\bf 1.} {\em The orbit-centered inertial frame
(OCI) $\{x_1,y_1,z_1\}$.} Here the $z_1$-axis is oriented along the
normal to the orbital plane of the satellite ($\sim 70.142\deg$
coinclination).  The $x_1$-axis is defined to be the intersection of
the orbital plane and the earth's equatorial plane, and the
origin is the center of mass of the earth.  We have assumed here that
this frame is inertial, and have not included the secular drag of the
line-of-nodes of the orbital plane due to the oblateness of the earth.
This $\sim126~\deg/yr$ precession can be included at the end of our
analysis as an adiabatic correction, and again has no qualitative
effects upon the dynamics.

{\bf 2.} {\em The earth-centered inertial (ECI) reference frame
$\{x_2,y_2,z_2\}$}. Here the $z_2$-axis is aligned with the body axis
of the earth. The $x_2$-axis lies in the earth's equatorial plane at
zero degrees longitude, and the origin coincides with the center of
the earth.

{\bf 3.} {\em The body frame (non-inertial) $\{x_b,y_b,z_b\}$.} The
origin is at the center of the satellite, and the axes correspond to a
set of principal axes.  The satellite is assumed to be a slightly
oblate ($\sim 3.27\%$) spheroid of brass, and the $z_b$-axis is
aligned along its body axis.  The $x_b$ and $y_b$ axes are an
arbitrary fixed set of orthogonal axes spanning the equatorial plane
of the satellite.  In our calculations, the body axis is related to
the orbit-centered frame through the three Euler angles $\theta$,
$\phi$, and $\psi$. The nutation angle $\theta$ is the angle between
$z_b$ and $z_1$, while the angle of precession $\phi$ is the angle
between the $x_1$-axis and the line of nodes (the intersection of the
orbital plane and the equatorial plane of the satellite ($x_b$-$y_b$
plane)).  The spin angle $\psi$ is the angle between the line of nodes
and the $x_b$-axis.

{\bf 4.} {\em The Landau-Lifshitz (non-inertial) coordinate system
$\{x^{ll}, y^{ll}, z^{ll}\}$.} The $z^{ll}$-axis is aligned along the
instantaneous angular momentum vector of the satellite ($\vec
\omega$). The $x^{ll}$-axis is picked so that the
instantaneous magnetic field ($\vec B$) at the satellite lies in the
$x^{ll}-z^{ll}$ plane. Note that $z^{ll}$ need not be aligned with the
body axis of the satellite, and in fact, during the asymptotic
behavior of the satellite they are vastly different.  In particular,
the angle ($\xi$) between the symmetry axis of the satellite ($\hat
z_b$) and the instantaneous angular momentum vector ($\hat z_{ll}$)
can be expressed in terms of the three Euler angles as
\begin{eqnarray}
\sin\xi = \left(1+{I_3^2\over I_1^2}{(\dot \phi\cos\theta+\dot\psi)^2
          \over (\dot\phi^2\sin^2\theta+\dot\theta^2)}\right)^{-1/2}
\label{*}
\end{eqnarray}
In the early stages of the LAGEOS mission when $\dot \psi \gg \dot
\theta$ and $\dot\psi\gg\dot\phi$ this angle is rather small ($\bar
\xi \approx (I_1/I_3) (\bar\omega_{\theta}/\omega_{\psi}) \approx
6.3516\times 10^{-10}$).
\bigskip

A proper interpretation of the results of the numerical simulation
requires a careful distinction between $\omega_p$ and $\omega_\phi$
(where $\phi$ is the Euler angle of the body frame).  The two coincide
only under the assumption that the $\omega_3$ component of the
satellite's angular velocity is responsible for all of the satellite's
energy, or, to put it differently, the angular momentum of the
satellite is directed along the body axis of the satellite. As the
satellite's spinning motion slows down, this last assumption is
violated.  $\omega_p$ represents only a part of $\omega_\phi$, the
other part being caused by the ``tidal locking'' effect. As we shall
see, this is exactly what happens in the asymptotic phase of the
satellite's motion.  Nevertheless, both $\omega_p$ and $\omega_\phi$
remain bounded.

\section{Spin Dynamics of LAGEOS: The Equations.}
\label{IV}
The spin dynamics are determined by Euler's equations
\begin{eqnarray}
I_1 \dot\omega_1 - \omega_2 \omega_3 (I_1 - I_3)&=& N_1, \nonumber \\
I_1 \dot\omega_2 - \omega_3 \omega_1 (I_3 - I_1)&=& N_2, \\
I_3 \dot\omega_3 &=& N_3, \nonumber
\label{6}
\end{eqnarray}
where $\omega_1$, $\omega_2$, $\omega_3$ are the components of the
satellite's angular velocity in the body frame, $I_1 = I_2$, $I_3$ are
the principal moments of inertia, and $N_1$, $N_2$, $N_3$ are the
components of the torques along the satellite's principal axes. After
substituting the expressions for $\omega_1$, $\omega_2$, $\omega_3$ in
terms of the Euler angles
\begin{eqnarray}
\omega_1 &=& \dot\phi \sin\theta \sin\psi + \dot\theta \cos\psi,
\nonumber \\
\omega_2 &=& \dot\phi \sin\theta \cos\psi - \dot\theta \sin\psi, \\
\omega_3 &=& \dot\phi \cos\theta + \dot\psi, \nonumber
\label{7}
\end{eqnarray}
the Euler equations become
\begin{eqnarray}
\ddot{\theta}&=& \ddot{\theta}_{free} + {N_1\cos{\psi}-
N_2\sin{\psi}\over I_1},\\
\ddot{\theta}_{free}&=&\left({I_1-I_3\over
I_1}\right)\dot\phi^2\cos{\theta}\sin{\theta}-{I_3\over
I_1}\dot\psi\dot\phi\sin\theta,\\
\ddot{\phi} &=& \ddot{\phi}_{free} +
{N_1\sin{\psi}+N_2\cos{\psi}\over I_1 \sin{\theta}},\\
\ddot{\phi}_{free}&=&\left({I_3-2I_1\over
I_1}\right){\cos{\theta}\over \sin{\theta}} \dot\theta\dot\phi +
{I_3\over I_1} {\dot\theta\dot\psi\over \sin{\theta}},\\
\ddot{\psi} &=& \ddot{\psi}_{free} + {N_3\over I_1} -
{N_1\sin{\psi}+N_2\cos{\psi}\over I_1} {\cos{\theta}\over
\sin{\theta}},\\
\ddot{\psi}_{free} &=& -\left({I_3-I_1\over I_1}\right)
{\cos^2{\theta}\over \sin{\theta}}\dot{\theta}\dot{\phi} +
{\dot{\theta}\dot{\phi}\over \sin{\theta}} - {I_3\over I_1}
{\cos{\theta}\over \sin{\theta}} \dot{\theta}\dot{\psi}.
\label{8}
\end{eqnarray}

The torque components $N_1$, $N_2$, $N_3$ are due to gravitational and
magnetic forces acting on the satellite
\begin{equation}
N_i = N_i^{(g)} + N_i^{(m)}, \quad i=1, 2, 3.
\label{9}
\end{equation}
Gravitational torques in the body frame are given by
\begin{eqnarray}
N_1^{(g)} &=& -\cos\psi {\partial V\over \partial\theta}
- {\sin\psi \over \sin\theta} {\partial V \over \partial\phi} +
{\cos\theta \sin\psi \over \sin\theta} {\partial V \over
\partial\psi},\nonumber \\
N_2^{(g)} &=& \sin\psi {\partial V\over \partial\theta}
- {\cos\psi \over \sin\theta} {\partial V \over \partial\phi} +
{\cos\theta \cos\psi \over \sin\theta} {\partial V \over
\partial\psi},\\
N_3^{(g)} &=& -{\partial V \over \partial\psi}. \nonumber
\label{10}
\end{eqnarray}
Using the standard dipole approximation, the gravitational potential $V$ is
\begin{equation}
V = {G M (I_3 -I_1) \over 2 R^3} \left( 3 \gamma^2 - 1\right),
\label{11}
\end{equation}
where $\gamma$ is the direction cosine between (1) the radial vector
from the satellite center of mass to the center of the earth, and (2)
the symmetry axis of the satellite. It is related to Euler's angles
via
\begin{equation}
\gamma = \sin\theta \sin (\eta - \eta_0 - \phi ),
\label{12}
\end{equation}
where $\eta$ gives the angular position of the center of mass of the
satellite in its orbit about the earth, and $\eta_0$ is an arbitrary starting
position.
The potential is given by
\begin{equation}
V = {G M (I_3 -I_1) \over 2 R^3} \left( 3 \sin^2\theta \sin^2(\eta -
\eta_0 -  \phi ) - 1\right),
\label{13}
\end{equation}
and thus
\begin{eqnarray}
-{\partial V\over\partial\theta} &=& -{3 G M (I_3 -I_1)\over R^3}
\sin\theta \cos\theta \sin^2(\eta - \eta_0 - \phi ), \nonumber \\
-{1\over\sin\theta} {\partial V\over\partial\phi} &=&
{3 G M (I_3 -I_1)\over R^3} \sin\theta \sin (\eta - \eta_0 - \phi )
\cos (\eta - \eta_0 - \phi ), \\
{\partial V\over\partial\psi} &=& 0.               \nonumber
\label{14}
\end{eqnarray}
Equations (10) and (14) lead to the following final expressions for the
components of the gravitational torque in the body frame
\begin{eqnarray}
N^{(g)}_1 &=& {3 G M (I_3 -I_1)\over R^3} \sin\theta
\sin (\eta - \eta_0 - \phi ) \left\{\begin{array}{c} -\cos\theta
\cos\psi \sin (\eta - \eta_0 - \phi ) \\
+ \sin\psi \cos (\eta - \eta_0 - \phi ) \end{array}\right\},\nonumber \\
N^{(g)}_2 &=& {3 G M (I_3 -I_1)\over R^3} \sin\theta
\sin (\eta - \eta_0 - \phi ) \left\{\begin{array}{c} \cos\theta \sin\psi
\sin (\eta - \eta_0 - \phi ) \\
+ \cos\psi \cos (\eta - \eta_0 - \phi ) \end{array}\right\}, \nonumber
\\ N^{(g)}_3 &=& 0.
\label{15}
\end{eqnarray}

As for the magnetic effects, we are interested in the torque
components acting on a conducting ball of radius $a$ spinning with
angular velocity $\vec\omega$ in an external magnetic field $\vec B$.
We take our expressions from Landau and Lifshitz \cite{LL}, noting
that their results are for what we have dubbed the Landau-Lifshitz
frame (as described above), and for our purposes need to be
transformed to the body frame.

As already described, the Landau-Lifshitz frame is determined by the
vectors $\vec\omega$ and $\vec B$.  If we introduce an arbitrary set
of rectangular coordinates $x$, $y$, $z$, in this frame $\vec\omega$
and $\vec B$ are represented as
\begin{eqnarray}
\vec\omega &=& \omega_x\, \hat x + \omega_y\, \hat y + \omega_z\, \hat
z =  \langle \omega_x, \omega_y, \omega_z\rangle, \\
\vec B &=& B_x\, \hat x + B_y\, \hat y + B_z\, \hat z =
\langle B_x, B_y, B_z\rangle,
\label{16}
\end{eqnarray}
where hats ($\hat x$) denote vectors normalized to unity.  The
transition between this arbitrary frame and the Landau-Lifshitz frame
is given by
\begin{equation}
\left(\begin{array}{c} \hat x^{ll}\\ \hat y^{ll}\\ \hat
z^{ll}\end{array} \right)
=\left(\begin{array}{ccc}[(\vec\omega\times\vec B)\times\vec\omega ]_x\over
\omega\vert\vec\omega\times\vec B \vert &
[(\vec\omega \times\vec B) \times \vec\omega ]_y\over \omega
\vert \vec\omega \times \vec B \vert &
[(\vec\omega \times\vec B) \times \vec\omega ]_z\over \omega
\vert \vec\omega \times \vec B \vert \\
(\vec\omega \times \vec B)_x\over \vert\vec\omega \times \vec B\vert &
(\vec\omega \times \vec B)_y\over \vert\vec\omega \times \vec B\vert &
(\vec\omega \times \vec B)_z\over \vert\vec\omega \times \vec B\vert \\
\omega_x\over\omega & \omega_y\over\omega & \omega_z\over\omega
 \end{array}\right)\,
\left(\begin{array}{c}\hat x \\ \hat y \\ \hat z
\end{array}\right), \label{17}
\end{equation}
where $\omega = \vert\vec\omega\vert$.

The components of the magnetic torque in the Landau-Lifshitz frame are
given by
\begin{eqnarray}
N^{ll}_x &=& V \alpha'' B^{ll}_x B^{ll}_z, \nonumber \\
N^{ll}_y &=& -V \alpha' B^{ll}_x B^{ll}_z, \\
N^{ll}_z &=& -V \alpha'' (B^{ll}_x)^2, \nonumber
\label{18}
\end{eqnarray}
where $V = 4 \pi a^3/3$ is the volume of the ball, $B^{ll}_x$,
$B^{ll}_y$, $B^{ll}_z$ are components of the magnetic field in
the Landau-Lifshitz frame, and the real and complex parts of the
coefficient of magnetization are
\begin{eqnarray}
\alpha' &=& -{3\over 8 \pi} \left[ 1 - {3\delta\over 2a} {{\rm sinh}
\left( 2 {a\over\delta}\right) - \sin \left( 2 {a\over\delta}\right)
\over {\rm cosh} \left( 2 {a\over\delta}\right) - \cos \left( 2
{a\over\delta}\right)} \right], \nonumber \\
\alpha'' &=& -{9 \delta^2\over 16 \pi a^2}
\left[ 1 - {a\over \delta} {{\rm sinh}
\left( 2 {a\over\delta}\right) + \sin \left( 2 {a\over\delta}\right)
\over {\rm cosh} \left( 2 {a\over\delta}\right) - \cos \left( 2
{a\over\delta}\right)} \right],
\nonumber
\label{19}
\end{eqnarray}
with
$$
\delta = {c\over\sqrt{2 \pi \sigma \omega}}.
$$
Here $c$ is the
speed of light and $\sigma$ is the specific conductivity of the
material forming the ball. At small values of $\omega$ ($\delta\gg
a$), $\alpha'$ and $\alpha''$ can be approximated by the expressions
\begin{eqnarray}
\alpha' &\approx & -{4 \pi\over 105} {a^4 \sigma^2 \omega^2 \over c^4},
\\
\alpha'' &\approx & {a^2 \sigma \omega \over 10 c^2}.
\label{20}
\end{eqnarray}

Components of the magnetic field can be evaluated in the ECI frame
using the dipole approximation. In the spherical coordinate
representation of the ECI frame, the magnetic field components are
\begin{eqnarray}
B_{r_1} &=& -{2M\over R^3} \cos\theta_1, \nonumber \\
B_{\theta_1} &=& -{M\over R^3} \sin\theta_1, \\
B_{\phi_1} &=& 0,  \nonumber
\label{21}
\end{eqnarray}
where $M$ is the magnetic dipole moment of the earth.  In the
rectangular coordinate representation of the ECI frame (the index 1 is
used everywhere for quantities in ECI)
\begin{eqnarray}
B_{x_1} &=& B_{r_1} \sin\theta_1 \cos\phi_1 + B_{\theta_1} \cos\theta_1
\cos\phi_1, \nonumber \\
B_{y_1} &=& B_{r_1} \sin\theta_1 \sin\phi_1 + B_{\theta_1} \cos\theta_1
\sin\phi_1, \\
B_{z_1} &=& B_{r_1} \cos\theta_1 - B_{\theta_1} \sin\theta_1. \nonumber
\label{22}
\end{eqnarray}
Transforming these to the OCI frame:
\begin{eqnarray}
B_{x_2} &=& (B_{r_1} \sin\theta_1 + B_{\theta_1} \cos\theta_1)
\cos\phi_1,    \nonumber \\
B_{y_2} &=& (B_{r_1} \sin\theta_1 + B_{\theta_1} \cos\theta_1) \sin\phi_1
\cos\xi + (B_{r_1} \cos\theta_1 - B_{\theta_1} \sin\theta_1), \sin\xi \\
B_{z_2} &=& -(B_{r_1} \sin\theta_1 + B_{\theta_1} \cos\theta_1) \sin\phi_1
\sin\xi + (B_{r_1} \cos\theta_1 - B_{\theta_1} \sin\theta_1) \cos\xi,
\nonumber
\label{23}
\end{eqnarray}
where $\xi$ is the colatitude angle of the orbit with respect to
magnetic north, and the index 2 is used for quantities in the OCI.
The satellite's semimajor axis is $R=1.227179\times 10^9~cm$, and for
the earth's magnetic dipole moment we used $M=7.9\times
10^{25}~G-cm^3$.  Although we do not average the magnetic field in our
simulations, it was useful for back of the envelope calculations to
note that the the magnetic field, averaged over one orbit, is $|\bar
B| = 0.155814~G$, with components,
\begin{eqnarray}
\bar B_{x_2} &=& 0~G, \nonumber \\
\bar B_{y_2} &=& -0.0388558~G,\\
\bar B_{z_2} &=& -0.150891~G. \nonumber
\label{501}
\end{eqnarray}

The angles $\theta_1$ and $\phi_1$ are dependent upon the satellite's
position in its orbit. The satellite's coordinates in the OCI frame are
\begin{eqnarray}
x_2 &=& r \cos (\eta - \eta_0), \nonumber \\ y_2 &=& r \sin (\eta -
\eta_0), \\ z_2 &=& 0, \nonumber
\label{24}
\end{eqnarray}
where
\begin{equation}
\eta = - {2 \pi \over T_{\rm orbit}} t.
\label{25}
\end{equation}
In ECI we have
\begin{eqnarray}
x_1 &=& R \cos (\eta - \eta_0), \nonumber\\
y_1 &=& R \sin (\eta - \eta_0) \cos\xi, \\
z_1 &=& - R \sin (\eta - \eta_0) \sin\xi. \nonumber
\label{26}
\end{eqnarray}
Hence,
\begin{eqnarray}
\sin\theta_1 &=& \sqrt{\cos^2(\eta - \eta_0) + \sin^2(\eta - \eta_0)
\cos^2\xi}, \\
\cos\theta_1 &=& - \sin (\eta - \eta_0) \sin\xi, \\
\cos\phi_1 &=& {\cos (\eta - \eta_0)\over \sin\theta_1}, \\
\sin\phi_1 &=& {\sin (\eta - \eta_0) \cos\xi \over \sin\theta_1}.
\label{27}
\end{eqnarray}

The three Euler equations (Eq.~\ref{8}), with the magnetic and
gravitational torques included (properly transformed to the body
frame), give us the vehicle to analyze qualitatively the spin dynamics
of the LAGEOS-1 satellite.  We present our results from the numerical
integration of these equations in the following two sections.
\bigskip

\section{Initial-Value Data.}
\label{V}
We solved the Euler equations (Eq.~\ref{8}) using a fourth-order
Bulirsch-Stoer algorithm with adaptive time-stepping \cite{NR}.  The
equations were integrated for $3\times 10^9$ seconds, as we wanted to
(1) reproduce the experimentally-measured spin rates ($17~yrs$ following
launch), (2) examine the spin-orbit resonance ($\sim 27~yrs$ after
launch), and (3) reveal the asymptotics of the spin dynamics ($\sim79+
yrs$ after launch). The experimentally-measured exponential decrease
in the spin rate imposed a constraint on our theoretical model, linking
the ``effective'' radius of the satellite ($a$) with the satellite's
``effective'' conductivity ($\sigma$):
\begin{equation}
\sigma a^5 \sim 1.19552\times 10^{24}~cm^5/s.
\label{28}
\end{equation}
The satellite was modeled as a $25.55~cm$ radius spheroid of brass
($\sigma = 1.098\times10^{17}~s^{-1}$).  LAGEOS I's moment of inertia
about the body axis is $I_{z_b}=1.314\times 10^8~g-cm^2$, while the
moment of inertia perpendicular to the body axis is $I_{x_b}=I_{y_b}=
1.271\times 10^8~g-cm^2$, corresponding to an approximately $3.38\%$
deviation from sphericity.

{}From the experimental data, we observed a deviation from pure
exponential damping of the spin of the satellite at early times. This
is presumably due to the satellite's transition from magnetic
opaqueness to transparency in the course of its spin damping.  A
rapidly rotating conductor, with angular velocity $\omega$ and
conductivity $\sigma$, will have an associated magnetic skin depth
\begin{eqnarray}
\delta = c/\sqrt{2 \pi \sigma \omega}.
\label{502}
\end{eqnarray}
This skin depth starts out considerably smaller than the satellite,
but as the satellite's spin is damped down, the skin depth becomes
much larger than the satellite's diameter.  Although this transition
effect will be more pronounced for our idealized spherical brass model
satellite than for LAGEOS (with its additional surface structure), it
helps account for the structure in the experimental spin-down data
(for which it would have been convenient to have access to the
associated error bars). It is for this reason that we began our
calculations within the transparent phase, at $t_0=92772864~s$. We
integrated forward in time to $t = 3\times 10^9~s$, and backwards to
$t=0~s$.  At $t_0$, the skin depth of our satellite model, given the
experimentally measured spin rate of $\omega_0=4.36332~s^{-1}$ and a
conductivity for brass ($\sigma=1.098\times 10^{17}~s^{-1}$), is $\sim
17.3~cm$.  We started the integration with the satellite in a
retrograde circular orbit, at a radius of $1.227179\times 10^9~cm$ and
an inclination of $109.859 ~\deg$.  At $t_0$, the satellite was
located over the equator and positioned along the $x_2=x_1$ axis
($\eta_0=0$).  We inserted the body axis of the satellite into the
orbital plane with the angular momentum vector parallel to the OCI
$y_1$-axis. The initial spin of the satellite was determined by the
experimental measurements, and was set to $4.36332~rad/sec$. The
initial conditions were
\begin{eqnarray}
 && \psi (t_0) = 0, \quad \theta (t_0) = 1.5707963, \quad \phi (t_0) =
0, \\
 && \dot\theta (t_0) = 0, \quad \dot\phi (t_0) = 0, \quad  \dot\psi
(t_0)=4.36332~s^{-1} .\nonumber
\label{29}
\end{eqnarray}
In running this initial data backward in time from $t_0=927728640~s$
to launch at $t=0$, we recovered the following angles and angular
velocities (measured in $s^{-1}$):
\begin{eqnarray}
 && \theta (t=0) = 1.304029, \quad \phi (0) = -1.327836, \quad \psi
(0) = -0.5542969 \\
 && \dot\theta (0) = 6.792913\times 10^{-10}, \quad \dot\phi (0) =
3.806309\times 10^{-9}, \quad \dot\psi (0) = 10.77009   . \nonumber
\label{29b}
\end{eqnarray}
The magnetic field in our simulation was assumed to be a perfect
dipole field, of moment $-M=7.9\times 10^{25}~G/cm^3$ and aligned along the
$z_2$-axis of the ECI frame.

\begin{figure}
\centerline{\epsfxsize=\figwidth\epsfbox{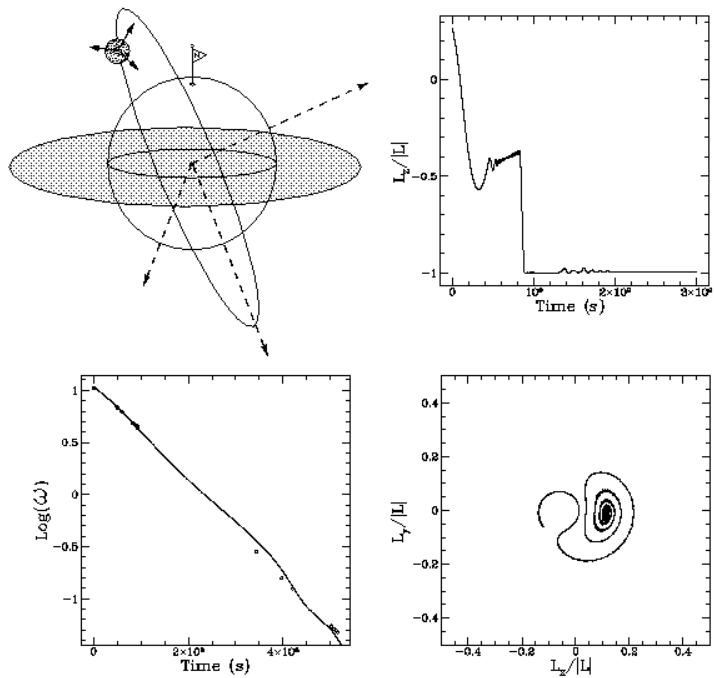}}
\caption{
{\bf The spin dynamics of LAGEOS: Numerical results}.  We present here
results from our numerical simulation of the dynamic evolution of the
LAGEOS satellite (here modeled as a slightly oblate spheroid of brass
orbiting an earth-like mass idealized as perfectly spherical and
endowed with a polar-oriented dipole magnetic field, top left figure).
The evolution of the satellite's angular momentum reveals three unique
phases, as demonstrated in the plot of the component of angular
momentum orthogonal to the orbital plane (top right). The first phase
is characterized by an exponential decrease in spin of the satellite,
with negligible nutation and precession.  The decay in spin for a
$25.55~cm$ radius sphere of brass agrees with the data (bottom left).
In the spin-orbit resonance phase the angular velocity decays to a
value comparable to the orbital angular velocity.  The nutation and
precession increase, and the angular momentum vector lifts off the
orbital plane and settles down orthogonal to the orbital plane (top
right). The third phase (asymptotic) is characterized by the tidal
locking of the satellite.}
\label{figure3}
\end{figure}

\section{Results.}
\label{VI}
We have identified three distinct phases in the spin dynamics of the
LAGEOS satellite \protect{(Figs.~3-5)}, which we
shall refer to as (1) the Fast Spin Phase, (2) the Spin-Orbit
Resonance Phase, and (3) the Asymptotic Phase.

a) {\em Fast Spin Phase}. This first phase, from $t=0$ to $t\sim
25~yrs$, is characterized by an exponential decrease in the spin rate
($\dot \psi$) of the satellite (Fig.\ \ref{figure3}) from $\dot
\psi(t=0)=10.7709~s^{-1}$ to $\dot \psi(t\sim 25~yrs)\approx.001~s^{-1}$.
The body axis of the satellite is aligned with the angular momentum of
the satellite in this phase; hence, all other quantities (angular
momentum, kinetic energy, {\em etc.}) also decrease exponentially.
The nutation of the satellite (the angle between the body axis of the
satellite and the normal to the orbital plane) undergoes a steady
increase from $\theta_0=90~\deg$ at $t=t_0$ to $\theta\sim 126~\deg$
over a period of $\sim 6.4~yrs$, indicating a nutation angular
velocity of $\bar \omega_{\theta} \approx 20~\deg/yr$. The nutation
then settles down to a ``pseudo-stable'' state at $\theta \approx
115~\deg$ over the next $3.2$ years and remains at this value ($\pm
5\%$) until the onset of the spin-orbit resonance at $\sim 25~yrs$
(Fig.\ \ref{figure6}).  Finally, within this twenty five year period
the satellite precesses in a positive sense by 63 revolutions before
unwinding when entering the spin-orbit resonance phase.

b) {\em Spin-Orbit Resonance Phase}. The spin dynamics abruptly
change at $t\sim 27~yrs$. This is precisely when the satellite's spin,
decreased by magnetic damping, approaches the orbital angular velocity
\begin{eqnarray}
\omega_{orbit} = \sqrt{{3.9\times 10^{20}\over (1.227179\times
10^9)^3}}~s^{-1} \approx 0.0004593782~s^{-1}.
\label{504}
\end{eqnarray}
The conductivity of the satelite was chosen to reproduce
experimentally observed exponential decay
in the spin rate of the satellite,
\begin{eqnarray}
\dot \psi(t) = \psi(t_0) e^{-{(t-t_0)\over \Delta}},
\label{505}
\end{eqnarray}
with ($\Delta \sim 2.96~yr$ time constant).  Therefore, the spin-orbit
resonance should occur at $t_{sp} \sim 30~yrs$, which is in agreement
with the numerical results. The resonance phase is marked by a
movement of the angular momentum vector to a position orthogonal to
the orbital plane, and is furthermore characterized by the beginning
of satellite wobble ({\em i.e.}, the point in time when $\xi$ of Eq.
(\ref{*}) becomes nonzero and the body axis becomes misaligned with
the instantaneous angular momentum vector). From this point forward it
is more illustrative to examine the dynamics of the instantaneous
angular velocity and momentum rather than the Euler angles. In
addition to the dramatic changes in the spin dynamics, this second
phase also gives rise to a reversal in the signs of the precessional
velocity ($\omega_{\phi}$) and spin ($\omega_{\psi}$).

c) {\em The Asymptotic Phase}.  Following the spin-orbit resonance
phase, the spin dynamics gradually settled down to an asymptotic
regime over the course of $\sim 50~yrs$.  Not surprisingly, the
satellite becomes tidally locked (think of the moon).  In particular,
the asymptotic value of the total angular velocity is equal to the
orbital angular velocity, subject to small fluctuations.  We note that
in this phase the torques induced from the changing magnetic field due
to {\em orbital} motion of the satellite will become important, and
should no longer be ignored. However, their addition will not
significantly change dynamics, as the energies at this point are quite
low.

\begin{figure}
\vspace{5cm}
\centerline{\epsfxsize=\figwidth\epsfbox{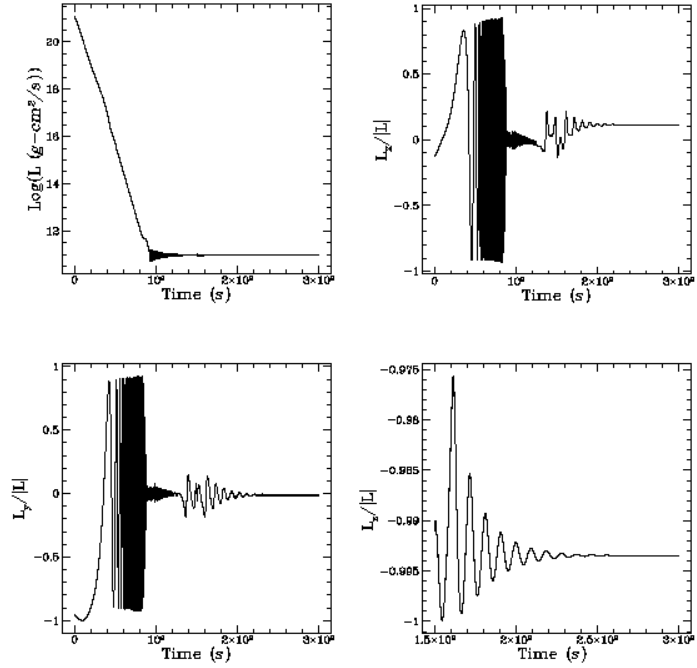}}
\caption{
{\bf The behavior of the angular momentum vector and its components
throughout the mission}. The evolution of the angular momentum reveals
the transition to an asymptotic phase wherein the angular momentum
vector is roughly orthogonal to the body axis of the satellite, and is
consistent with a precession rate equal to the orbital angular
velocity.}
\label{figure4}
\end{figure}

{}From our numerical runs, a rough estimate of the asymptotic behavior
of the satelite model (modulo phase and a finite offset in $\phi$) is
given by
\begin{eqnarray}
\theta(t) &\simeq & 1.57+0.12\cos(4.6\times 10^{-4} t), \\
\phi(t) &\simeq & -4.59\times 10^{-4} t,\\
\psi(t) &\simeq & 3.04+0.115\cos(4.6\times 10^{-4} t),\\
\omega_{\theta}(t)  &\simeq &  5.5\times 10^{-5}\cos(4.6\times 10^{-4}
t),\\
\omega_{\phi}(t) &\simeq & -4.59\times 10^{-4}+3.1\times
10^{-3}\cos(1.4\times 10^{-4} t),\\
\omega_{\psi}(t) &\simeq & 5.3\times 10^{-5}\cos(4.6\times 10^{-4} t).
\label{505b}
\end{eqnarray}
The asymptotic values of the other relevant parameters (Fig.\ \ref{figure4}):
\begin{eqnarray}
K.E. &\sim& 13.42 ergs,\\
-{1\over c}(M\dot B) &\sim& 3.2\times 10^{-8} ergs,\\
G.E. &\sim& -0.45375 ergs \ to \ -0.45 ergs,\\
E &\sim& 12.96 ergs,\\
L &\sim& 5.84\times 10^4 g\,cm^2/s,\\
\hat L_x &\sim& 0.115,\\
\hat L_y &\sim& -0.15 \pm 0.05,\\
\hat L_z &\sim& -0.9933,\\
\omega &\sim& 0.0004595 s^{-1}
\label{506}
\end{eqnarray}

\begin{figure}
\vspace{5cm}
\centerline{\epsfxsize=\figwidth\epsfbox{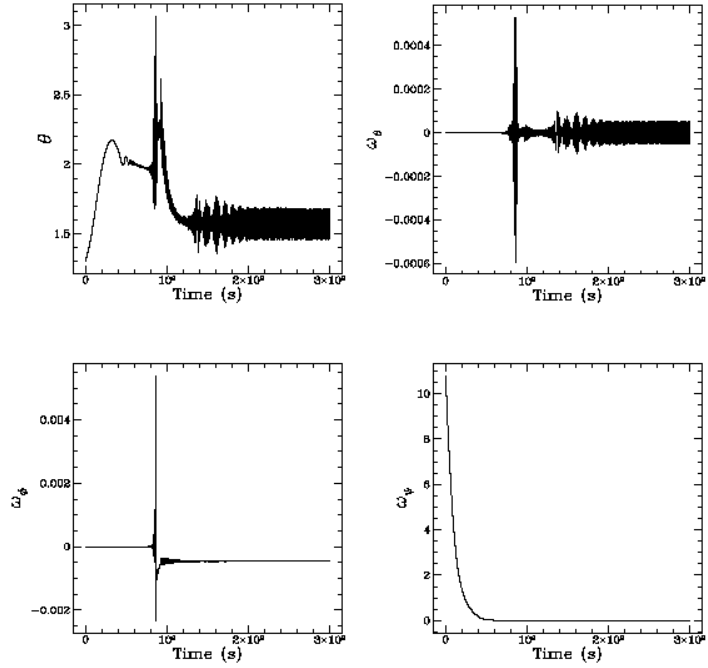}}
\caption{
{\bf Dynamics of the Euler angles}. The evolution of the nutation
(angle of obliquity, $\theta$) of the model satellite (upper left). In
the asymptotic limit the angular velocities of nutation and precession
average to zero (upper and lower right, respectively). The precession
rate ($\omega_{\phi}$), on the other hand, locks into the orbital
velocity (lower left). This last plot demonstrates clearly the
dynamics through the spin-orbit resonance phase.}
\label{figure5}
\end{figure}

\begin{figure}
\vspace{5cm}
\centerline{\epsfxsize=\figwidth\epsfbox{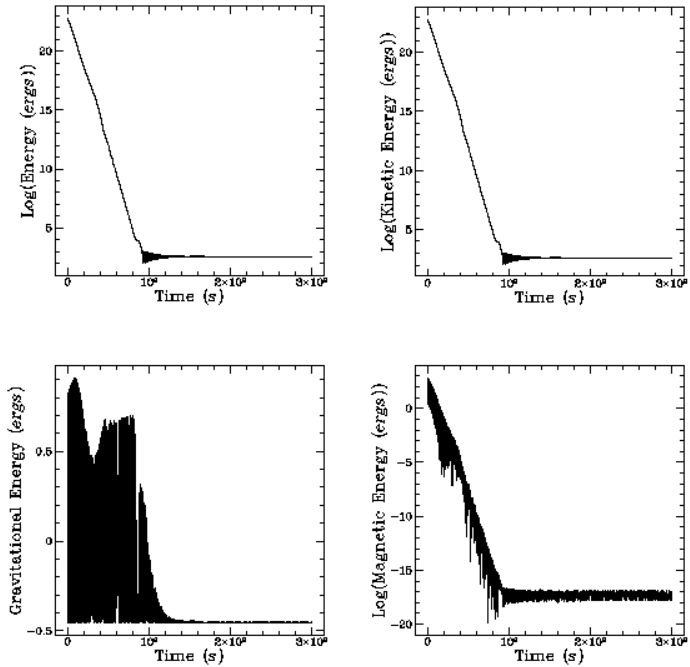}}
\caption{
{\bf Energy}. A plot of the logarithm of the total energy of the
satellite as a function of time reveals an asymptotic limit of
$12.96~ergs$ consistent with tidal locking. The linear region early on
($0-9\times 10^8~s$) reflects the resistive damping phase of the
induced eddy currents. Asymptotically, the satellite orients itself so
as to minimize the gravitational potential energy (lower left) while
the magnetic energy (${1\over c}M\cdot B$) becomes negligible.}
\label{figure6}
\end{figure}

\section{Conclusions.}
\label{VII}
Of the five largest sources of error identified in the LAGEOS-3
experiment, the earth and solar-induced surface forces are potentially
the most troublesome (Table\ \ref{table1}). The anisotropic heating of the
satellite, and subsequent reradiation, gives a ``thermal rocketing''
perturbation (referred to as the Rubincam effect or the Yarkowsky
thermal drag) which tends to degrade the experiment. To model this
effect requires, in part, a detailed knowledge of the behavior of the
angular momentum of the satellite.  Toward this end we have derived,
and solved numerically, a simplified set of Euler equations that
evolve the angular momentum vector for a slightly oblate spheroid of
brass orbiting an earth-like mass, idealized as being a perfect sphere
and having a perfect polar-oriented dipole magnetic field. The Euler
equations included both the tidal gravitational torques and the
eddy-current torques, as well as the resistive damping torques, as
modeled by complex magnetization coefficients. Using this rather
simplified model, we have identified three phases of the rotational
dynamics -- a fast spin phase, a spin-orbit resonance phase, and an
asymptotic phase (Fig.\ \ref{figure3}). We have also identified an error in the
previously established model of asymptotic spin dynamics~\cite{BI}.
This error has led to confusion and, in attempts to reconcile observed
data with theoretical predictions, has led others to hypothesize
erroneous models for the moments of inertia of LAGEOS-1 \cite{SCHAR}.

Our results have led us to formulate four as yet unresolved questions:
(1) Can we obtain the asymptotic solution analytically, and in so
doing can we understand the wobbling or slippage of the Euler angles
with respect to the relatively stable total instantaneous angular
velocity?; (2) Can we understand why the {\em rms} fluctuations in the
gravitational potential energy cause the asymptotic value of the
angular momentum vector to be offset from the orbital plane by $\sim
10~deg$?; (3) Can we understand why the nutation angle ($\theta$)
drifts initially at a rate of $\sim 20~deg/yr$ and reaches a
pseudo-stable value of $\sim 115~deg$?; and (4) Can we understand the
fluctuations in the spin rate ($\dot \psi$) over the first $\sim
20~yrs$, which do not appear to have been detected experimentally?  We
are addressing these questions by (1) introducing a more realistic
model of the satellite and earth into our calculations \cite{HALV},
and (2) exploring more of phase space by way of Poincare sections. The
results presented here provide us with clues that must be pieced
together to reveal the physics behind the complex motion we observe.

The current spin dynamics model suggests that we launch LAGEOS-3, with
as large a spin $(\dot\psi)$ as possible, into an obliquity of $\theta\sim
115~\deg$; although, due to the qualitative nature of our results,
the precise numbers are far from being conclusive.

We are currently working with the Center for Space Research at the
University of Texas at Austin to determine the impact this revised
model of the spin dynamics of LAGEOS will have on the LAGEOS-3 mission
(in particular, how will the Rubincam effect alter the line of nodes
of the orbital plane?). In addition, we are working closely with
colleagues at the University of Texas and the University of Maryland
to reconcile the experimental measurements of the spin dynamics of
LAGEOS-1 with our theoretical model \cite{LAGEXP}. Furthermore, we
will propose an optimal experimental measurement schedule in support
of the proposed LAGEOS-3 mission.

\acknowledgements
We wish to thank Stirling Colgate, Douglas Currie, Christopher Fuchs,
and Sara Matzner for many helpful discussions.  This work was
supported in part by a grant from Los Alamos National Laboratory under
LDRD XL31, by the AFOSR under the Summer Faculty Research
Program, and by NSF grants PHY~88-06567 and PHY~93-10083.

\end{document}